\documentclass[aps,prb,reprint,superscriptaddress]{revtex4-1}

\usepackage{graphicx}
\usepackage{amssymb,amsmath}
\usepackage{bm,color}

\begin{document}

\title{Minimal phase-field crystal modeling of vapor-liquid-solid coexistence and transitions}

\author{Zi-Le Wang}
\affiliation{State Key Laboratory of Low Dimensional Quantum Physics, Department of Physics, Tsinghua University,
Beijing 100084, China}
\author{Zhirong Liu}
\email{LiuZhiRong@pku.edu.cn}
\affiliation{College of Chemistry and Molecular Engineering, and Beijing National Laboratory for
  Molecular Sciences (BNLMS), Peking University, Beijing 100871, China}
\author{Zhi-Feng Huang}
\email{huang@wayne.edu}
\affiliation{Department of Physics and Astronomy, Wayne State University, Detroit, Michigan 48201, USA}
\author{Wenhui Duan}
\email{duanw@tsinghua.edu.cn}
\affiliation{State Key Laboratory of Low Dimensional Quantum Physics, Department of Physics,
  Tsinghua University, Beijing 100084, China}
\affiliation{Institute for Advanced Study, Tsinghua University, Beijing 100084, China}
\affiliation{Collaborative Innovation Center of Quantum Matter, Tsinghua University, Beijing 100084, China}
\affiliation{Frontier Science Center for Quantum Information, Beijing 100084, China}

\date{\today}

\begin{abstract}

A new phase field crystal model based on the density-field approach incorporating high-order
interparticle direct correlations is developed to study vapor-liquid-solid coexistence and
transitions within a single continuum description. Conditions for the realization of phase
coexistence and transition sequence are systematically analyzed, and shown to be satisfied by
a broad range of model parameters, demonstrating the high flexibility and applicability of the
model. Both temperature-density and temperature-pressure phase diagrams are identified, while
structural evolution and coexistence among the three phases are examined through dynamical
simulations. The model is also able to produce some temperature and pressure related material
properties, including effects of thermal expansion and pressure on equilibrium lattice spacing,
and temperature dependence of saturation vapor pressure. This model can be used as an effective
approach for investigating a variety of material growth and deposition processes based on
vapor-solid, liquid-solid, and vapor-liquid-solid growth.

\end{abstract}

\maketitle

\section{Introduction}
\label{sec: introduction}

The vapor-based growth techniques, such as chemical vapor deposition (CVD), vapor-phase epitaxy
(VPE), physical vapor deposition (PVD), and vapor-liquid-solid (VLS) growth, have been widely
adopted in the fabrication and synthesis of two-dimensional (2D) and three-dimensional (3D)
thin film materials \cite{re:stangl04,AquaPhysRep13,Yazyev14}, heterostructures
\cite{LiuAdvMater18,NanoLett993}, and nanowires \cite{HannonNature06,afm323}. The interaction
between vapor and solid or liquid phases plays an important role during the growth process
since it determines the interfacial morphology and microstructures (including the formation
of topological defects such as dislocations and grain boundaries) which affect the mechanical,
electrical, magnetic, and thermal properties of the sample. A comprehensive understanding
of the detailed dynamical process and underlying mechanisms, which are key in achieving
high-quality material systems, is a challenging task for both experimental \textit{in situ}
studies and computer simulations given the multiple spatial and temporal scales involved.
Both atomistic and coarse-grained modeling and simulation methods have been developed and
applied to the study of these complex growth dynamics and mechanisms. For example, molecular
dynamics simulations can probe into atomic-level microstructural details of the CVD
\cite{jpcc6097,npj14} and VLS \cite{Haxhimali09,NC1956} growth processes. However, they
are usually limited by the small simulation time scales (around ns to $\mu$s) and system
sizes that are far from reaching those of real experimental systems. Another widely used
modeling technique is the phase field method \cite{Annu113,Annu163,re:wang10}, which is a
coarse-grained, mesoscale approach at the long-wavelength limit, with the capability of
describing system evolution on diffusion time scales including that of interfacial morphology
in CVD and VLS growth \cite{Schwalbach11,Cryst2211,prm033402,jpcc9902}. Despite its advantage
on accessing large length and time scales, phase field models are short of the description
of short-wavelength, microscopic scales such as crystalline details and defect microstructures,
and need to incorporate additional elastic, plastic, or orientation fields to account for the
effects of elastoplasticity, defects, and multiple grain orientations.

Given its unique capacity in combining atomic-scale spatial resolution with diffusive
time-scale dynamics and its intrinsic incorporation of elastoplasticity and multiple
orientations, the phase field crystal (PFC) method \cite{prl245701,pre51605,prb064107}
has been developed rapidly in recent years as a useful tool in studying a wide range
of phenomena of materials growth, structural evolution, and transformation. Its
applications involve many important physical processes such as solidification
\cite{pre51605,prb064107,pre051404,pre031603,Taha19}, thin film epitaxy
\cite{re:huang08,re:wu09}, crystal growth \cite{prl035702,pre012405}, dynamics of
dislocations \cite{prl015502,Berry14,SkaugenPRB18,SkaugenPRL18,Salvalaglio20} and grain
boundaries \cite{olmsted11,prl255501,SalvalaglioPRMater18,Zhou19}, and the formation
of quasicrystals \cite{prl255501_2,prl075501} and heterostructures \cite{prb165412}.
Most of early PFC models were constructed based on two-point correlation to describe
systems governed by isotropic interactions \cite{prb064107}, where the crystal
structures and ordered patterns are controlled by microscopic lattice length scales
\cite{prl045702,jpc205402,pre53305,prl35501,prl205502}. Limited work has been
attempted to explore the influence of orientation-dependent interactions and
higher-order correlations \cite{jpc364102,prm060801,prb35447}. Recently, we have
developed an angle-adjustable PFC formulation to provide a complete and concise way
to incorporate any $n$-point correlations for modeling crystalline systems that are
rotationally invariant and governed by both isotropic and anisotropic interparticle
interactions \cite{prb180102}. From this approach various 3D and 2D crystalline
structures (such as bcc, simple cubic, diamond cubic, simple monoclinic, orthorhombic,
hexagonal, rhombic, and square phases) have been simulated. Such a complete density-field
formulation further expands the scope of PFC models in the study of a variety of complex
phase behaviors, and will be the basis of model development in this work.

A limitation of most PFC models is that the modeling is usually restricted
to liquid and solid phases and the related transition processes, but not involving
vapor phase and its coupling or coexistence with solid or liquid state that are
essential in simulating the widely used growth processes (e.g., CVD, VPE, PVD,
and VLS growth) for the synthesis of thin films and nanostructures. Schwalbach
{\it et al.}~made the first attempt to incorporate vapor phase into the PFC method
\cite{pre023306}, which requires an extra order parameter field (in addition to the
PFC density field) in the free energy functional to generate realistic liquid-vapor
and vapor-solid interfacial properties and step-flow growth. By assuming the
long-wavelength approximation of three- and four-point correlations, Kocher and
Provatas developed another PFC model with the use of a single PFC density field to
effectively model vapor-liquid-solid transitions and simulate the growth processes
involving two or three phases \cite{prl15501}. The model has been
extended to incorporate the coupling to thermal transport \cite{KocherPRM19}, and
the pressure control dynamics introduced in the model has been further developed
and applied to the study of binary alloy systems \cite{prm083404}.

In this paper we present a new and efficient vapor-liquid-solid PFC model based on
the general density-field approach, with the expansion of three- and four-point direct
correlations in terms of gradient nonlinearities in the free energy functional. The
model efficiency can be viewed from its relatively simple form, serving as a minimal
theory for modeling vapor-liquid-solid coexistence and transitions.
The advantage of the model can be also seen from its
tunability in achieving three-phase coexistence and the desired transition sequence
across a broad range of model parameter values. The conditions and properties of
these phase coexistence and transitions are calculated analytically and numerically,
and verified through 2D dynamical simulations. In addition, we demonstrate the ability
of this model in obtaining realistic material properties of, e.g., saturation vapor
pressure, thermal expansion and pressure-induced contraction of crystalline lattice
spacing which are absent in the existing PFC models.
Since this model is built on the density-field formulation of Ref.~\cite{prb180102} with
a universal formalism for the expansion of any $n$-point correlations satisfying the
condition of rotational invariance, it can be readily extended to incorporate 
bond-angle dependent anisotropic interactions (as in Ref.~\cite{prb180102}) into the
three-phase formulation constructed here, to simulate a broader category of material systems.

\section{Model}
\label{sec:model}

In the original PFC model, the free energy functional is given by \cite{prl245701,pre51605,prb064107}
\begin{equation}
\mathcal{F}\left[n(\mathbf{r})\right]=
\int{d\mathbf{r}\left [ -\frac{n}{2} \left( C_0+C_2\nabla ^2+C_4\nabla ^4 \right) n
  -\frac{E_0}{4!}n^4 \right ]},
\label{F_2p_corr} 
\end{equation}
where $n(\mathbf{r})$ denotes the order parameter field of atomic number density variation, 
and the parameters $C_0$, $C_2$ and $C_4$ can be connected to the two-point direct correlation
function in classical density functional theory \cite{prb064107}. To enable the description
of a spatially periodic, crystalline phase, $C_2<0$ and $C_4<0$ are required. Also $E_0<0$ is
needed to prevent the divergence of density fluctuation. Via rescaling the length and time
scales \cite{pre51605}, Eq.~(\ref{F_2p_corr}) can be converted into the simplest form of
\begin{equation}
\mathcal{F}\left[n(\mathbf{r})\right]
=\int{d\mathbf{r}\left\{ \frac{1}{2}n \left[-\epsilon+(\nabla^2+1)^2 \right] n
+\frac{1}{4}n^4 \right\}},
\label{F_2pcorr}
\end{equation}
where the only remaining parameter $\epsilon$ reflects the influence of the temperature. 
The larger the $\epsilon$ value, the lower the temperature it corresponds to.

The original PFC Eq.~(\ref{F_2pcorr}) contains only two-point direct correlation and excludes
proper vapor-liquid-solid transitions. To incorporate the contributions from three- and four-point
correlations, we adopt the general density-field approach developed in Ref.~\cite{prb180102}
which formulates the condition of rotational invariance in the expansion of any order of direct
correlation functions, and consider the following minimal form of the free energy functional
\begin{eqnarray}
 && \mathcal{F}[n(\mathbf{r})] = -\int B_0 n(\mathbf{r})d\mathbf{r} \nonumber\\
 && \quad -\frac{1}{2} \int n(\mathbf{r})\left(C_0+C_2\nabla^2 + C_4\nabla^4 + C_6\nabla^6 \right)
 n(\mathbf{r}) d\mathbf{r} \nonumber\\
 && \quad  -\frac{1}{3!} \int \left[D_0 n^3(\mathbf{r}) + D_{11}n^2(\mathbf{r})\nabla^2 n(\mathbf{r})
 \right] d\mathbf{r} \nonumber\\
 && \quad  -\frac{1}{4!} \int \left\{E_0n^4(\mathbf{r})
 +E_{1122}n^2(\mathbf{r})\left[\nabla^2 n(\mathbf{r})\right]^2\right\} d\mathbf{r}, 
 \label{Functional}
\end{eqnarray}
where $E_0<0$ and $E_{1122}\leq 0$. The linear term with coefficient $B_0$ was usually
ignored in most PFC studies since its integration over space gives a constant
proportional to the average density $\bar n$ and thus does not change the relative
stability among different phases and the system dynamics. However, it was
demonstrated recently that this term is crucial for the calculation and control of system
pressure and elastic constants \cite{prb144112}. As will be shown below, $B_0$ should be
temperature dependent to give a correct behavior of saturation vapor pressure. In this model
parameters $C_0$, $C_2$, and $C_4$ also depend on temperature. Terms $D_{11}n^2\nabla^2 n$ and
$E_{1122}n^2(\nabla^2 n)^2$, corresponding to the contributions from three- and four-point
correlation respectively, are the main new components of our model and are key to achieve
the coexistence and transitions between vapor, liquid, and solid phases, as will be proved
both analytically and numerically in the next sections. A negative $E_{1122}$ is required
in the presence of $D_{11}$ to prevent the free-energy divergence of ordered phases, as
will be explained in Sec.~\ref{sec:VLS}. In addition, the $C_6$ term is introduced
to better control the crystalline modes in the presence of those two new nonlinear
gradient terms, but not essential for obtaining the vapor-liquid-solid transitions.
It is important to note that
in contrast to previous PFC models \cite{pre51605,prb064107,prl35501,prl205502},
here contributions from two-point correlation alone [i.e., $C_{j=0, 2, 4, 6}$ terms in
Eq.~(\ref{Functional})] are not enough to determine even the lowest-order structural
properties. The three- and four-order interactions play an important role in this new model,
as can be seen in, e.g., the corresponding homogeneous-state structure factor derived in
Appendix \ref{sec:Sq}. 

Although in principle more higher-order rotationally invariant terms from three- and
four-point correlations can be introduced through the formulation of Ref.~\cite{prb180102}, 
Eq.~(\ref{Functional}) is sufficient to produce vapor-liquid-solid transitions and serves
as the corresponding minimal PFC model when considering only isotropic interactions.
This model is convenient to be implemented, analyzed, and extended, with an important
feature being that the coexistence of three phases and the triple point can be realized
across a relatively broad range of parameters, as will be demonstrated below.
In addition to its simpler form, the model is constructed with the use
of the mere condition of rotational invariance, as compared to the previous two versions
of PFC models incorporating vapor-liquid-solid phases \cite{pre023306,prl15501} which
rely on some specific pre-assumptions of free energy terms or interparticle correlation
functions. In the model of Ref.~\cite{pre023306} by Schwalbach {\it et al.}, an additional
order parameter field and the associated free energy functional were needed for the
control of vapor phase; the model of Ref.~\cite{prl15501} by Kocher and Provatas also
made use of three- and four-point direct correlation functions, while assuming them as
products of Gaussian-type functions in Fourier space that correspond to infinite series
of nonlinear gradient terms in real space. Importantly, the new model introduced here
can capture some fundamental material properties, such as thermal expansion and some
pressure-related effects which are important in the modeling of real material systems
but are absent in these previous PFC models. Detailed analyses of our model will be
given in the next section.

\section{Analysis of vapor-liquid-solid transitions}
\label{sec:analysis}

\subsection{Vapor-liquid coexistence}
\label{sec:VL}

Both vapor and liquid are uniform phases with constant but different values
of density $n(\mathbf{r})=\bar n$, where $\bar n$ is the average density variation of
the system. Substituting it into Eq.~(\ref{Functional}) yields a simple Landau free energy
per volume
\begin{equation}
  f_{\rm u}(\bar n)=-B_0\bar n-\frac{1}{2}C_0 \bar n^2-\frac{1}{6}D_0 \bar n^3
  - \frac{1}{24} E_0 \bar n^4, \label{V-L_Eq1}
\end{equation}
with a single variable $\bar n$.

\begin{figure*}
  \centerline{\includegraphics[width=0.9\textwidth]{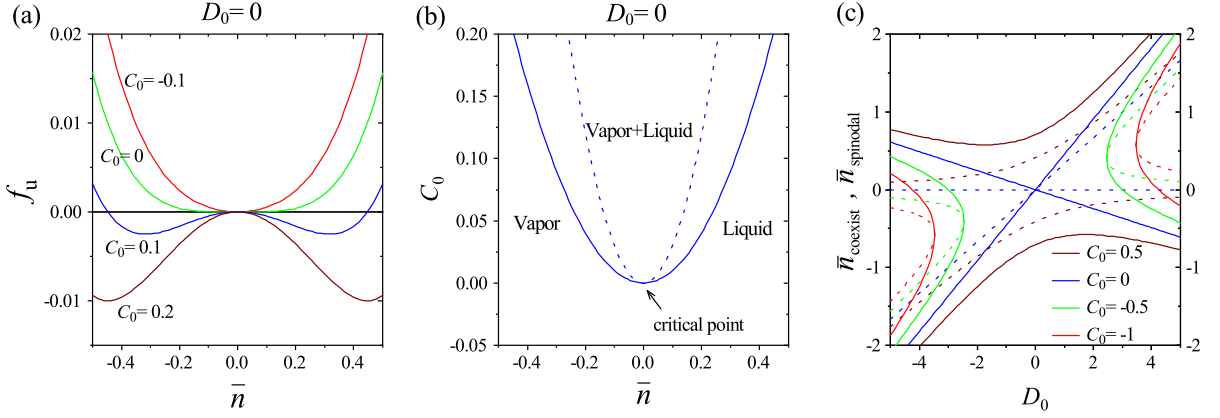}}
  \caption{Coexistence of vapor and liquid phases if not considering the solid state. 
    (a) The free energy density of uniform phase as a function of $\bar n$ at different $C_0$
    ($=\epsilon-1$). (b) The $C_0$-$\bar n$ phase diagram for vapor and liquid states, where
    the boundary of vapor-liquid coexistence is plotted as solid lines and the spinodal line
    is plotted as dashed. $D_0=0$ is set in both (a) and (b). (c) The density of vapor-liquid
    coexistence ($\bar n_{\rm coexist}$, solid curves) or the spinodal density ($\bar n_{\rm spinodal}$,
    dashed curves) as a function of $D_0$ at different values of $C_0$. Other parameters are
    $B_0=0$ and $E_0=-6$.}
\label{fig:Fig1}
\end{figure*}

In the rescaled original PFC model Eq.~(\ref{F_2pcorr}), we have $B_0=D_0=0$,
$C_0=\epsilon-1$,  and the temperature-related parameter $\epsilon$ is usually assumed
to be small \cite{pre51605,prb064107}. In that case, the $f_{\rm u}(\bar n)$ curve is
convex, and thus there exists only one single phase under any density or pressure. 
In other words, vapor-liquid coexistence is absent under small $\epsilon$. However,
when we extend the parameter region to large $\epsilon$, vapor-liquid coexistence
actually occurs based on Eq.~(\ref{V-L_Eq1}) without considering the solid state.
Some sample curves of $f_{\rm u}(\bar n)$ near $C_0=0$ are plotted in Fig.~\ref{fig:Fig1}(a),  
which shows a double-well free energy when $C_0$ is positive, i.e., in the regime of
$\epsilon>1$. The corresponding phase diagram with vapor-liquid coexistence is given
in Fig.~\ref{fig:Fig1}(b), where the critical point locates at $\bar n=0$, $C_0=0$
($\epsilon=1$). Below the critical point, vapor and liquid are indistinguishable and
there is no vapor-liquid coexistence. Above the critical point, vapor-liquid coexistence
occurs and the coexistence regime expands with increasing $C_0$. Note that this result
applies in the absence of solid phase which could become more stable in this parameter
regime in the original PFC model.

The vapor-liquid coexistence is affected by the cubic term with nonzero $D_0$. Applying the
common tangent rule on $f_{\rm u}(\bar n)$ in Eq.~(\ref{V-L_Eq1}), values of $\bar n$ for the
vapor and liquid phases in coexistence are given by 
\begin{equation}
\bar n_{\rm coexist}=\frac{-D_0\pm \sqrt{3D_{0}^{2}-6C_0E_0}}{E_0}.
\label{ncoexist}
\end{equation}
The spinodal densities are determined by $\partial^2 f_{\rm u}/\partial {\bar n}^2=0$, yielding
\begin{equation}
\bar n_{\rm spinodal}=\frac{-D_0\pm \sqrt{D_{0}^{2}-2C_0E_0}}{E_0}.
\label{nspinodal}
\end{equation}
The corresponding results are plotted in Fig.~\ref{fig:Fig1}(c) as a function of $D_0$. 
Introducing $D_0$ expands the region of vapor-liquid coexistence. 
For example, in the absence of $D_0$ the coexistence occurs only when $C_0>0$; in contrast,
as shown in Fig.~\ref{fig:Fig1}(c) with $E_0=-6$, at $D_0=\pm 3.46$ the coexistence
regions is expanded to $C_0>-1$, i.e., to smaller values of $\epsilon$ within the scope of
original PFC model.

In short, the above analysis indicates that the vapor-liquid coexistence can be realized
within the PFC framework of a single density order parameter when $D_{0}^{2}>2C_0E_0$
(with $E_0<0$), under either large enough $C_0$ (or $\epsilon$) or large enough $|D_0|$.

\subsection{Conditions for vapor-liquid-solid transitions}
\label{sec:VLS}

\begin{figure*}
  \centerline{\includegraphics[width=0.9\textwidth]{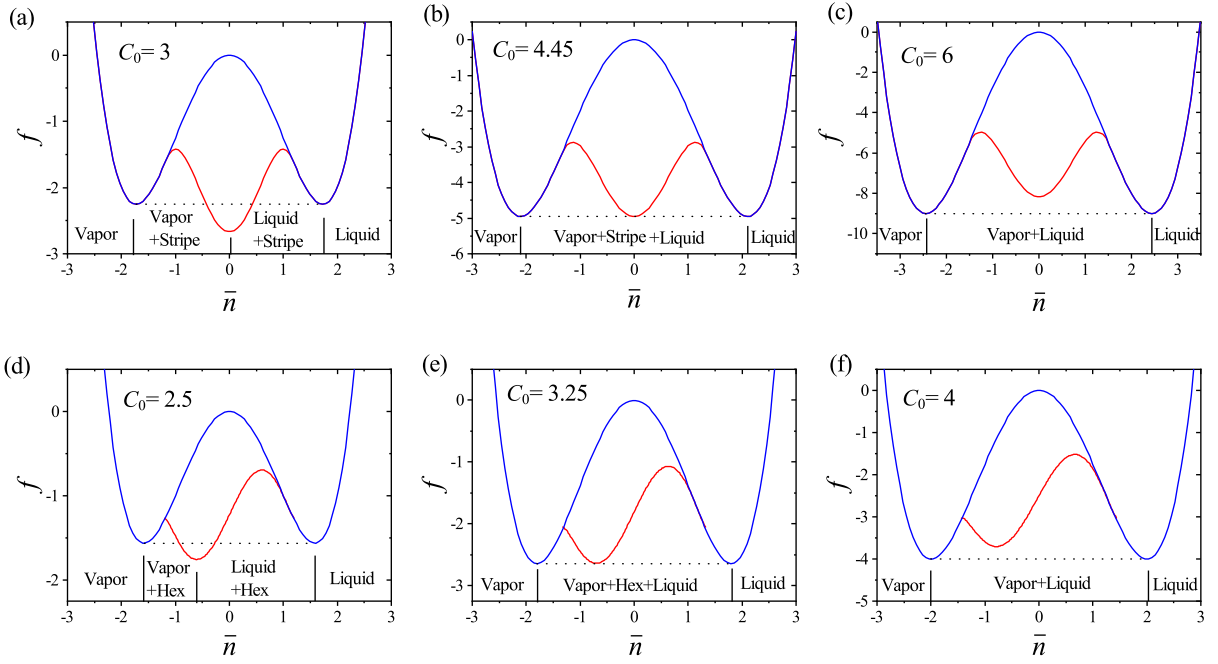}}
  \caption{Equilibrium free energy density of uniform (blue) and solid (red) phases as a function
    of $\bar n$ at different values of $C_0$ in one-mode approximation, where the solid phase is
    of (a)--(c) 1D stripe or (d)--(f) 2D hexagonal (Hex) symmetry. The parameters used are
    $B_0=C_6=D_0=D_{11}=E_{1122}=0$, $C_2=-2$, $C_4=-1$, and $E_0=-6$.}
\label{fig:Fig2}
\end{figure*}

With the knowledge of vapor-liquid coexistence given above, we now explore the way to realize
vapor-liquid-solid transitions. To simplify the problem and facilitate theoretical analysis,
we adopt a one-mode approximation for $n(\mathbf{r})$ of periodic solid phases. 
For a one-dimensional (1D) stripe phase with amplitude $A$ and wave number $q$,
\begin{equation}
  n(\mathbf{r})=\bar n +A\left(e^{iqx}+\textrm{c.c.}\right).
  \label{n_Stripes}
\end{equation}
where ``c.c.'' represents complex conjugate.
Substituting it into Eq.~(\ref{Functional}) yields the corresponding free energy density
\begin{eqnarray}
  && f_{\rm stripe}\left(q,A;\bar n \right) 
  =-B_0\bar n-\frac{1}{2}C_0 \bar n^2-\frac{1}{6}D_0 \bar n^3 - \frac{1}{24} E_0 \bar n^4 \nonumber \\
  &&\quad -\left[\left(C_0-C_2q^2+C_4q^4\right)+\frac{1}{3}\left(3D_0-2D_{11}q^2\right)\bar n 
    \right. \nonumber\\
  &&\quad \left. +\frac{1}{12}\left(6E_0+E_{1122}q^4\right)\bar n^2\right]A^2
  -\frac{1}{4}\left(E_0+E_{1122}q^4\right)A^4. \nonumber\\
  \label{F_Stripes}
\end{eqnarray}
Note that for simplicity, here we assume $C_6=0$ in the free energy as the presence of $C_6$ term
would not affect the basics of vapor-liquid-solid transition sequence. The specific role played
by nonzero $C_6$ will be discussed separately at the beginning of Sec.~\ref{sec:phasediagram}.
Similarly, for a 2D hexagonal or triangular phase the density field is expanded as
\begin{equation}
  n(\mathbf{r})=\bar n +A\sum_{\mathbf{q}}\left(e^{i\mathbf{q}\cdot\mathbf{r}}+\textrm{c.c.}\right),
\end{equation}
where the basic wave vectors $\mathbf{q}=q(1,0)$, $q(1/2,\sqrt{3}/2)$, and $q(1/2,-\sqrt{3}/2)$.
The free energy density is then written by
\begin{eqnarray}
&& f_{\rm hex}\left(q,A;\bar n \right)
  =-B_0\bar n-\frac{1}{2}C_0 \bar n^2-\frac{1}{6}D_0 \bar n^3 - \frac{1}{24} E_0 \bar n^4 \nonumber \\
  && \quad -3\left[\left(C_0-C_2q^2+C_4q^4\right)+\frac{1}{3}\left(3D_0-2D_{11}q^2\right)\bar n 
    \right. \nonumber\\
  && \qquad\quad \left.+\frac{1}{12}\left(6E_0+E_{1122}q^4\right)\bar n^2\right]A^2 \nonumber\\
  && \quad -\left[2\left(D_0-D_{11}q^2\right)+\left(2E_0+E_{1122}q^4\right)\bar n\right]A^3 \nonumber \\
  && \quad -\frac{15}{4}\left(E_0+E_{1122}q^4\right)A^4. \label{F_Rods}
\end{eqnarray}
The equilibrium state is determined by minimizing the free energy density, i.e., 
$\min_{q,A} f(q,A;\bar n)$ via $\partial f / \partial q =0$ and
$\partial f / \partial A =0$. The corresponding results of equilibrium free energy density
for stripe and hexagonal phases are plotted in Fig.~\ref{fig:Fig2} as a function of $\bar n$.

The first row of Eq.~(\ref{F_Stripes}) or Eq.~(\ref{F_Rods}) is identical to that of the
uniform phase, either liquid or vapor, given in Eq.~(\ref{V-L_Eq1}). To identify the conditions
for achieving the vapor-liquid-solid transition sequence as $\bar n$ increases, in the following
we consider the linear instability of the supercooled or supersaturated uniform phase with
respect to the formation of the crystalline state, which is determined by the $A^2$ term 
in Eqs.~(\ref{F_Stripes}) and (\ref{F_Rods}), both being proportional to
\begin{eqnarray}
\alpha(q,\bar n) = - && \left[\left(C_0-C_2q^2+C_4q^4\right)+\frac{1}{3}\left(3D_0-2D_{11}q^2\right)
  \bar n \right. \nonumber\\
&& \left. +\frac{1}{12}\left(6E_0+E_{1122}q^4\right)\bar n^2\right].
\label{VLS_termA2}
\end{eqnarray}
When $\alpha<0$, a uniform liquid or vapor phase is linearly unstable under infinitesimal
fluctuations and will crystallize spontaneously. To find the minimum of $\alpha$ with respect
to $q$, we solve $\partial \alpha /\partial q=0$, leading to
\begin{equation}
q^2=\left\{
\begin{array}{cc}
\frac{6C_2+4D_{11}\bar n}{12C_4+E_{1122}{\bar n}^2}, & {\rm if~} 3C_2+2D_{11}\bar n <0,\\
0, & {\rm if~} 3C_2+2D_{11}\bar n \geq 0,
\end{array} \right.
\label{VLS_Eq2}
\end{equation}
where $12C_4+E_{1122}{\bar n}^2<0$ is required to prevent the divergence at large $q$ as can be
obtained from Eq.~(\ref{VLS_termA2}). Substituting Eq.~(\ref{VLS_Eq2}) into Eq.~(\ref{VLS_termA2})
yields
\begin{equation}
  \alpha(\bar n)= -\left(C_0+D_0\bar n+\frac{1}{2}E_0\bar n^2\right)
  +\frac{\left(3C_2+2D_{11}\bar n\right)^2}{3\left(12C_4+E_{1122}{\bar n}^2\right)},
  \label{VLS_Eq3}
\end{equation}
for $3C_2+2D_{11}\bar n <0$.

If $D_{11}=E_{1122}=0$ as in the original PFC model, we have $q^2={C_2}/{2C_4}$ which is independent
of $\bar n$, and $C_2<0$ is needed to enable solid phases. Defining the supercooling or
supersaturating density for the occurrence of linear instability by $\alpha(\bar n)=0$, we have
\begin{equation}
\bar n_{\rm supercool}^{({\rm o})}=\frac{-D_0\pm \sqrt{D_{0}^{2} -2E_0 \left(C_0-{C_2^2}/{4C_4}\right)}}{E_0}.
\label{nsupercool}
\end{equation}
A solid phase would be more stable than a uniform phase (vapor or liquid) when $\bar n$
lies in between the two values of $\bar n^{({\rm o})}_{\rm supercool}$. Comparing Eq.~(\ref{nsupercool})
with Eqs.~(\ref{ncoexist}) and (\ref{nspinodal}), it is clear that the midpoint of two
$\bar n^{({\rm o})}_{\rm supercool}$ coincides with that of $\bar n_{\rm coexist}$ or
$\bar n_{\rm spinodal}$ for vapor-liquid phases.
Therefore, the stability regime of solid phase is expected to locate in between those of vapor
and liquid, and the phase transition sequence is thus vapor-solid-liquid with the increase of
density, consistent with the results of free energy density curves given in Fig.~\ref{fig:Fig2}
for both 1D and 2D systems. The vapor-solid-liquid coexistence (corresponding to the triple point
in phase diagram) can be realized via adjusting model parameters appropriately, as shown in
Figs.~\ref{fig:Fig2}(b) and \ref{fig:Fig2}(e). In this case, the density of solid is smaller
than that of liquid, mimicking the unusual property of ice vs.~water but not the behavior of most
other materials. The above analysis hence demonstrates that it is impossible to describe the
usual vapor-liquid-solid transition sequence in the original PFC model with $D_{11}=E_{1122}=0$.

\begin{figure*}
  \centerline{\includegraphics[width=0.7\textwidth]{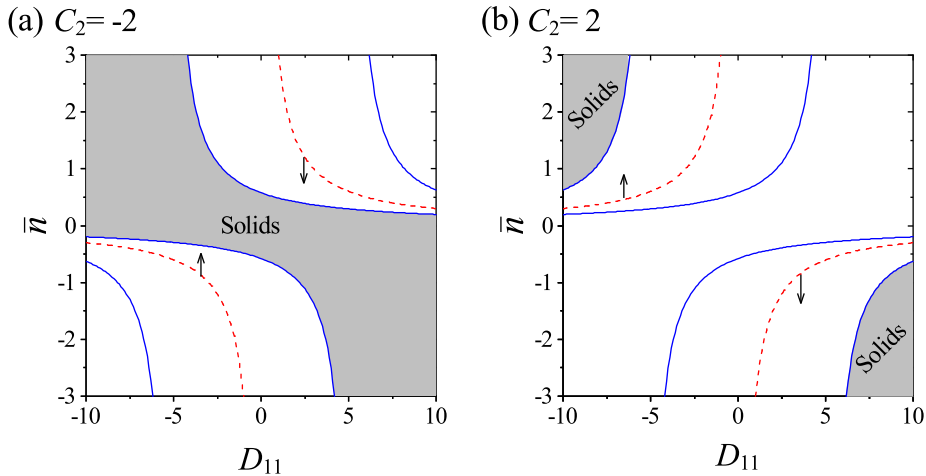}}
  \caption{Variation of $\bar n_{\rm supercool}$ with $D_{11}$. Calculations are based on
    Eq.~(\ref{nsupercool-2}) shown as solid curves. The dashed lines correspond to
    $3C_2+2D_{11}\bar{n}=0$, while arrows point to the region of $3C_2+2D_{11}\bar{n}<0$. 
    Regions for solid phase are shown in shadow. The parameters used are
    $B_0=C_6=D_0=E_{1122}=C_0=0$, $C_4=-1$, $E_0=-6$, and (a) $C_2=-2$, (b) $C_2=2$.}
\label{fig:Fig3}
\end{figure*} 

To achieve the usual vapor-liquid-solid transition sequence, we need to consider the effect
of nonzero $D_{11}$. When $E_{1122}=0$, $\alpha(\bar n)$ in Eq.~(\ref{VLS_Eq3}) keeps its
quadratic form of $\bar n$ and the linearly unstable condition of $\alpha(\bar n)=0$ is still
solvable analytically, yielding
\begin{widetext}
\begin{equation}
\bar n_{\rm supercool}
=\frac{
-\left(D_0-\frac{C_2 D_{11}}{3C_4}\right)
\pm \sqrt{
\left(D_0-\frac{C_2 D_{11}}{3C_4}\right)^2 
-2 \left( E_0-\frac{2D_{11}^2}{9C_4} \right) \left(C_0-\frac{C_2^2}{4C_4}\right)
}
}
{E_0-\frac{2D_{11}^2}{9C_4}}.
\label{nsupercool-2}
\end{equation}
\end{widetext}
Now the midpoint of two $\bar n_{\rm supercool}$ values does not coincide with that of
$\bar n_{\rm coexist}$ or $\bar n_{\rm spinodal}$ anymore. More importantly, when the quadratic
coefficient of $\alpha(\bar n)$ in Eq.~(\ref{VLS_Eq3}), i.e., $-{E_0}/{2} +{D_{11}^2}/{9C_4}$,
is negative, the solid phase is more stable than the vapor or liquid uniform phase for
$\bar n$ lying outside the range confined by the two $\bar n_{\rm supercool}$ values, but not
in between them as before. This makes it possible to tune the phase stability parameters
such that the density of solid would be higher than that of liquid, i.e., to realize the
usual vapor-liquid-solid transition sequence. Some examples are illustrated in Fig.~\ref{fig:Fig3},
where $C_0$ is set as 0 to approach the vapor-liquid coexistence, and $D_0=0$ is assigned
by properly choosing the reference state so that vapor and liquid phases locate at opposite
sides of $\bar n=0$ in the parameter space. In such a case, a vapor-liquid-solid transition
sequence requires the solid phase to be on the positive side of $\bar n$. However, for
$C_2=-2$ as used in most PFC models, the stability regime for solid always contains the
point $\bar n=0$ [see Fig.~\ref{fig:Fig3}(a)]; i.e., at $\bar n=0$ the solid state is more
stable than the uniform phase. This can be easily verified from Eqs.~(\ref{VLS_Eq2}) and
(\ref{VLS_Eq3}) which show that $q^2>0$ and $\alpha(\bar n)<0$ always hold at $\bar n=0$ for any
$C_0\ge 0$ (which is necessary for vapor-liquid coexistence when $D_0=0$; see Sec.~\ref{sec:VL}),
$C_2<0$, and $C_4<0$. It is noted that based on Eq.~(\ref{VLS_Eq2}), $3C_2+2D_{11}\bar{n}<0$
is needed for the appearance of solid state, giving $C_2<0$ in the absence of $D_{11}$ as in
previous PFC models. Conversely, with the introducing of nonzero $D_{11}$, $C_2<0$ is no
longer obligatory (see also Appendix \ref{sec:Sq}). When $C_2=2$ [Fig.~\ref{fig:Fig3}(b)],
in the $D_{11}$-$\bar n$ diagram the stability regime for solid phase shrinks as compared
to the case of $C_2=-2$, and locates at large enough $|D_{11}|$. Importantly, the solid
phase is not stable near $\bar{n}=0$, leaving space for vapor-liquid coexistence to occur. 
As seen in Fig.~\ref{fig:Fig3}(b), for small enough negative $D_{11}$ the density of solid
is higher than that of uniform (vapor or liquid) phase as desired.

\begin{figure}
  \centerline{\includegraphics[width=0.4\textwidth]{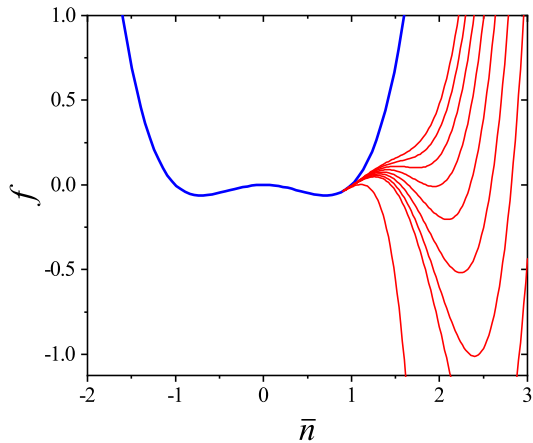}}
  \caption{Equilibrium free energy density of uniform (blue) and stripe (red) phases as a
    function of $\bar n$ in one-mode approximation, for various values of $E_{1122}$. 
    Result of uniform phase is not affected by $E_{1122}$, while for stripes the value of
    free energy density increases with the decrease of $E_{1122}$, i.e., $E_{1122}=0, -0.22,
    -0.24, -0.26, -0.28, -0.30, -0.32, -0.34, -0.36$ (red curves, from bottom to top).
    Other parameters are $B_0=C_6=D_0=0$, $C_0=0.5$, $C_2=2$, $C_4=-1$, $E_0=-6$, and
    $D_{11}=-8$.}
  \label{fig:Fig4}
\end{figure} 

The existence of proper vapor-liquid-solid transitions also requires the contribution of
the $E_{1122}$ term. The reason is that although nonzero $D_{11}$ enables the stabilization
of solid phase at density $\bar n$ larger than that of vapor and liquid phases, 
it overstabilizes the solid phase at very large $\bar n$. Take the stripe phase as an example, 
for which the free energy density at $E_{1122}=0$ is
\begin{eqnarray}
  && f_{\rm stripe}(\bar n)=-B_0\bar n-\frac{1}{2}C_0 \bar n^2-\frac{1}{6}D_0 \bar n^3
  - \frac{1}{24} E_0 \bar n^4 \nonumber \\
  && + \frac{1}{E_0} \left[\frac{(3C_2+2D_{11}\bar n)^2}{18C_4}-\left(C_0+D_0\bar{n}
    +\frac{1}{2}E_0\bar{n}^2 \right)\right]^ 2,
\label{F_Stripes2}
\end{eqnarray}
when $3C_2+2D_{11}\bar{n}<0$. The value of $f_{\rm stripe}(\bar n)$ is dominated by the $\bar{n}^4$
terms when $\bar{n} \gg 1$. To prevent $f_{\rm stripe} \rightarrow -\infty$, it is
required that
\begin{equation}
- \frac{1}{24} E_0 +\frac{1}{E_0} \left[ \frac{2D^2_{11}}{9C_4}-\frac{E_0}{2} \right ]^2 >0,
\label{E_1111_IN}
\end{equation}
which however is incompatible with the condition of $-E_0/2 + D_{11}^2/9C_4<0$ for the occurrence
of vapor-liquid-solid transition sequence as discussed above. Therefore, a negative $E_{1122}$ is
necessary to remedy this, as demonstrated in Fig.~\ref{fig:Fig4} which shows the increase of
$f_{\rm stripe}$ and the avoidance of divergence as $E_{1122}$ becomes more negative. 

In the next section we will conduct numerical calculations beyond one-mode approximation to
achieve the three-phase coexistence and transition, based on the above theoretical analyses
and the conditions identified for the realization of proper vapor-liquid-solid transitions.

\section{Numerical results}
\label{sec:Numerical}

Numerical calculations are conducted through the use of the time-evolution equation
\begin{equation}
\frac{\partial n}{\partial t}=\nabla^2 \frac{\delta\mathcal{F}[n]}{\delta n},
\label{DynamEq}
\end{equation}
which describes the conserved dynamics of density variation field $n(\mathbf{r},t)$.
Given Eq.~(\ref{Functional}) for the free energy functional $\mathcal{F}$ of this model,
the above dynamic equation is of the explicit form
\begin{eqnarray}
  \frac{\partial n}{\partial t} &&= \nabla^2 \Biglb \{ - \left (C_0 + C_2 \nabla^2 + C_4 \nabla^4
    + C_6 \nabla^6 \right ) n \nonumber\\
  && -\frac{1}{2} D_0 n^2 - \frac{1}{6} D_{11} \left ( 2n\nabla^2 n + \nabla^2 n^2 \right )
     \nonumber\\
  && -\frac{1}{6} E_0 n^3 -\frac{1}{12} E_{1122}
     \left [ n (\nabla^2 n)^2 + \nabla^2 (n^2 \nabla^2 n) \right ] \Bigrb \}.    
\end{eqnarray}
It is essentially governed by the diffusive, relaxational dynamics, and the system free energy
decreases with time $t$ continuously until it reaches an equilibrium or steady state.

\subsection{Vapor-liquid-solid coexistence and phase diagrams}
\label{sec:phasediagram}

\begin{figure}
  \centerline{\includegraphics[width=0.4\textwidth]{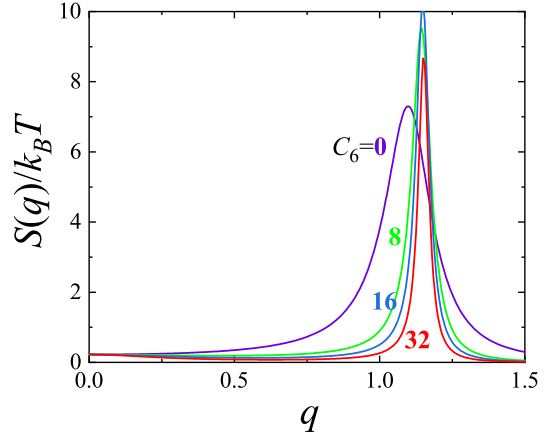}}
  \caption{The fluid-state structure factor $S(q)$ at different values of $C_6$, evaluated from the
    analytic result derived in Appendix \ref{sec:Sq}. The curves are plotted with four sets of
    parameters, including $(C_6, C_2, C_4)=(0, -10.6, -2.88)$, $(8, 3.63, 18.5)$, $(16, 17.8, 39.8)$,
    and $(32, 46.3, 82.5)$, with other parameters $C_0=-5.75$, $D_0=-9$, $E_0=-6$, $D_{11}=-34.2$,
    $E_{1122}=-52.1$, and $\bar{n}=-0.15$ remaining the same for each set. Each parameter set
    would lead to a state of vapor-liquid-solid coexistence under one-mode approximation of
    2D hexagonal structure when $\bar{n}_{\rm vapor}=-2.5$, $\bar{n}_{\rm liquid}=-0.5$,
    $\bar{n}_{\rm solid}=0$, and $A=0.2$. As $C_6$ increases the peak position of $S(q)$
    approaches the value $q=2/\sqrt{3}$ used in one-mode approximation.}
\label{fig:Fig5}
\end{figure} 

Our above analyses have demonstrated that the free-energy functional Eq.~(\ref{Functional}) with
$C_6=0$ is sufficient in obtaining the vapor-liquid-solid transitions under one-mode approximation. 
However, when solving the full PFC model via e.g., the dynamical Eq.~(\ref{DynamEq}), higher-order
modes play a non-neglectable role and could cause undesired disturbances on the phase behavior. 
To enhance the dynamical stability of the one-mode-like solutions, we introduce the nonzero $C_6$
term into the two-point direct correlation, which can be used to control the degree of
contributions from high-order modes on system properties. 
An example is given in Fig.~\ref{fig:Fig5}, showing some sample results of equilibrium fluid-state
structure factor $S(q)$ (as derived in Appendix \ref{sec:Sq}) for different values of $C_6$, each
of which corresponds to a set of model parameters giving vapor-liquid-solid coexistence. These
results indicate that contributions from high-order modes can be effectively suppressed at large
$C_6$.
 
\begin{figure}
  \centerline{\includegraphics[width=0.5\textwidth]{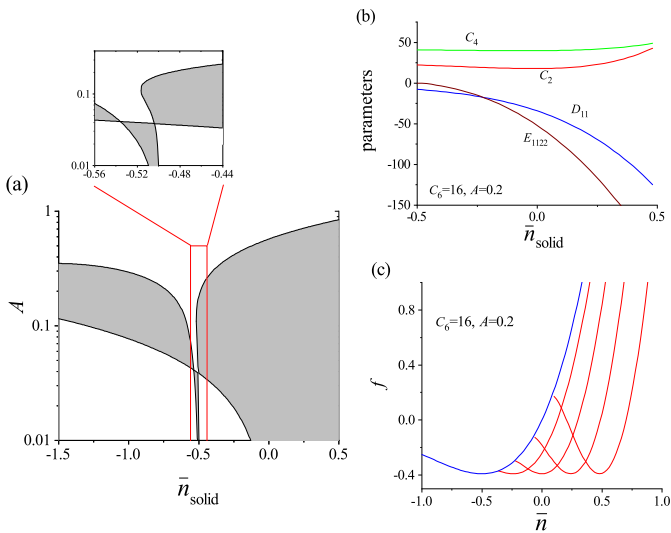}}
  \caption{The broad range of model parameters yielding three-phase coexistence, under the
    condition of fixed values of $B_0=-1.875$, $C_0=-5.75$, $D_0=-9$, and $E_0=-6$ such that
    vapor-liquid coexistence occurs at $\bar{n}_{\rm vapor}=-2.5$ and $\bar{n}_{\rm liquid}=-0.5$.
    (a) Values of solid-phase coexistence density $\bar{n}_{\rm solid}$ and one-mode amplitude
    $A$ for 2D hexagonal phase that can lead to existence of solutions for vapor-liquid-solid
    or vapor-solid-liquid coexistence [across all possible combinations of $(C_2, C_4, D_{11},
    E_{1122})$], as indicated by the shaded region. The results are generated for
    $C_6=16$, with very similar outcomes for other choices of $C_6>0$.
    (b) The allowed values of parameter set $(C_2, C_4, D_{11}, E_{1122})$ to achieve three-phase
    coexistence at different $\bar{n}_{\rm solid}$ when $C_6=16$ and $A=0.2$. 
    (c) The free energy density curves of liquid (blue) and solid (red) phases corresponding
    to four of the parameter sets in (b) that give $\bar{n}_{\rm solid}=-0.25, 0, 0.25, 0.5$,
    respectively. The vapor-phase free energy density (not shown here) is minimized at
    $\bar{n}=-2.5$ that forms a common tangent with these liquid- and solid-phase curves.
    The procedure of calculations under one-mode approximation is given in Appendix \ref{sec:Params}.
  }\label{fig:Fig6}
\end{figure} 

With the introduction of nonzero $C_6$, we can identify a broad range of parameters that lead to 
vapor-liquid-solid coexistence in the new PFC model developed here. The general procedure for
identifying the corresponding model parameters are described in Appendix \ref{sec:Params}, which
needs to be combined with some analytic conditions derived above in Sec. \ref{sec:analysis}
in the absence of $C_6$ (particularly $D_0^2>2C_0E_0$, $D_{11}<0$, and $E_{1122}<0$). Numerical
calculations are needed even in one-mode approximation, with some results presented in
Fig.~\ref{fig:Fig6}. Without loss of generality, in this example we fix the parameters
$C_0=-5.75$, $D_0=-9$, and $E_0=-6$ so that vapor-liquid coexistence is found at
$\bar{n}_{\rm vapor}=-2.5$ and $\bar{n}_{\rm liquid}=-0.5$ from Eq.~(\ref{ncoexist}).
We then search for all the possible values of $C_2$, $C_4$, $D_{11}$, and $E_{1122}$ that
give the solution of three-phase coexistence. Results in Fig.~\ref{fig:Fig6}(a) indicates
that the solution exists across a broad range of solid-phase coexistence density
$\bar{n}_{\rm solid}$ and amplitude $A$. [It is interesting to note that in addition to
vapor-liquid-solid coexistence (with $\bar{n}_{\rm vapor} < \bar{n}_{\rm liquid} 
< \bar{n}_{\rm solid}$), the parameter range for the unusual vapor-solid-liquid coexistence
(with $\bar{n}_{\rm vapor} < \bar{n}_{\rm solid} < \bar{n}_{\rm liquid}$) can also be identified in
this model, as seen in the part of $-1.5 \leq \bar{n}_{\rm solid} < -0.5$ in Fig.~\ref{fig:Fig6}(a).]
In other words, at any specific $\bar{n}_{\rm solid}$ within this range the associated
values of parameter set $(C_2, C_4, D_{11}, E_{1122})$ can be found to achieve three-phase
coexistence [see Fig.~\ref{fig:Fig6}(b)]. This continuous adjustability of model parameters is
demonstrated in an example of Fig.~\ref{fig:Fig6}(c), where $\bar{n}_{\rm solid}$ is pre-selected
from $-0.25$ to $0.50$ and for each of them we can always identify the corresponding combination
values of model parameters [given in Fig.~\ref{fig:Fig6}(b)] to obtain vapor-liquid-solid
coexistence.

\begin{figure}
  \centerline{\includegraphics[width=0.4\textwidth]{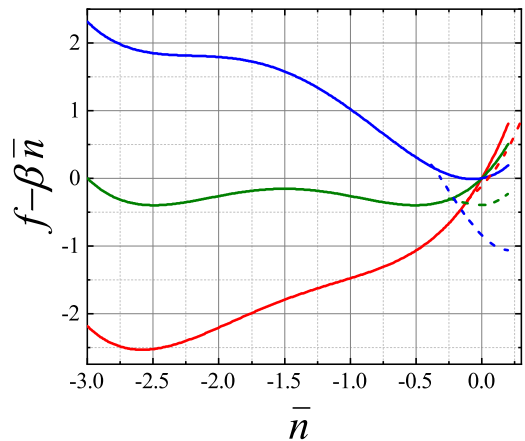}}
  \caption{Free energy density profiles of uniform (solid curves) and 2D hexagonal (dashed)
    phases at different temperatures $\Delta T = -0.514$ (blue), $\Delta T = 0$ (green), and
    $\Delta T = 0.486$ (red), using model parameters listed in Table~\ref{tab:table1}.
    Results are obtained from numerical solution of the full dynamical Eq.~(\ref{DynamEq}).
    Here all the free energy density curves have been tilted by a factor of $-\beta \bar n$ with
    $\beta=2.596$ for a better illustration.}
\label{fig:Fig7}
\end{figure}

\begin{table*}
  \caption{\label{tab:table1}Model parameters used in numerical calculations of vapor-liquid-solid
    transitions. $\Delta T=0$ corresponds to the triple point temperature.}
\begin{ruledtabular}
\begin{tabular}{ccccccccc}
  $B_0$   & $C_0$    & $C_2$    & $C_4$   & $C_6$ & $D_0$ & $D_{11}$ & $E_0$ & $E_{1122}$ \\ \hline
  $-4.5-3\Delta T$ & $-5.764-\Delta T$ & $17.8+2\Delta T$ & $39.8-\Delta T$ & $16$ & $-9$
  & $-34.2$ & $-6$ & $-52.1$\\
\end{tabular}
\end{ruledtabular}
\end{table*}

To obtain accurate values of the solid-phase equilibrium free energy beyond one-mode approximation,
we have numerically solved the full dynamical Eq.~(\ref{DynamEq}) using a single unit cell with
periodic boundary conditions, and calculated the free energy density of its equilibrium, steady
state. The initial density field $n(\mathbf{r},t=0)$ is set up either from the one-mode solution 
or from the existing simulation result of close parameter values. In addition, the numerical grid
spacings $\Delta x$ and $\Delta y$ are varied to determine the equilibrium wave number and thus
lattice constant from the minimum point of the corresponding free energy density obtained from
simulations at each $\bar n$ and temperature. The resulting equilibrium free energy density for
solid phase is then lower than that of one-mode approximation (although by a very small degree
due to the effect of nonzero $C_6$ term), and we can slightly adjust the model parameters to
achieve the desired phase stability and coexistence.

All the model parameters identified and used in the following full-model numerical calculations
are summarized in Table~\ref{tab:table1}, where $C_0$, $C_2$, and $C_4$ are set to be dependent
on an effective temperature $\Delta T$ for the coexistence among vapor, liquid, and solid phases
(with $\Delta T = 0$ being the triple point temperature). Parameter $B_0$ for the linear term
of the free energy functional is also set as temperature dependent, to produce the proper property
of pressure (see below). Examples of the resulting equilibrium profiles of free energy density
$f(\bar n)$ are given in Fig.~\ref{fig:Fig7}, at three different effective temperatures. At low
temperature ($\Delta T=-0.514$) both vapor-liquid and vapor-solid coexistence can be identified
from the $f$-$\bar n$ curves through the common tangent construction, while increasing temperature
to $\Delta T = 0$ brings the system to a vapor-liquid-solid coexistence as determined by the
common tangent of the vapor, liquid, and solid free energy curves. Further increasing the
temperature ($\Delta T = 0.486$) excludes vapor-solid coexistence while the separate vapor-liquid
and liquid-solid coexistence still remains. At high enough temperature, only liquid-solid
coexistence can be found. All these results are consistent with the well-known behavior of
the three phases.  

\begin{figure}
\centerline{\includegraphics[width=0.35\textwidth]{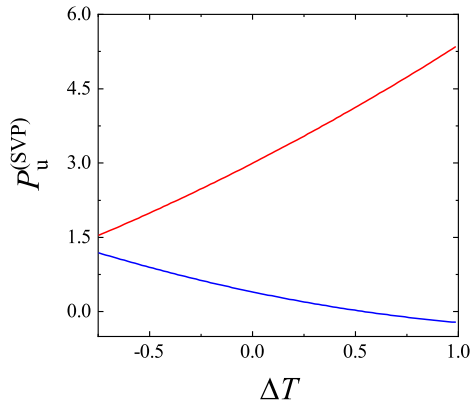}}
\caption{Saturation vapor pressure at vapor-liquid coexistence as a function of temperature
  $\Delta T$. Upper red line: $B_0=-4.5-3\Delta T$ as set in Table~\ref{tab:table1};
  Bottom blue line: $B_0=0$.}
\label{fig:Fig8}
\end{figure}

The pressure $P$ is also determined by $f(\bar n)$. In the PFC approach, the quantitative
result of pressure depends on the physical interpretation of the density variation field $n$
used in the model \cite{prb144112}. Here we adopt the interpretation that $n=(\rho-\rho_0)/\rho_0$,
where $\rho$ is the atomic number density and $\rho_0$ is a reference-state density. The total
number of particles $N$ in the system is kept constant under any deformations of volume $V$,
with $N=\int \rho d\mathbf{r} =\bar{\rho}V=\rho_0(\bar{n}+1)V$ where $\bar{\rho}$ is the spatial
average of $\rho(\mathbf{r})$, leading to ${\partial \bar{n}}/{\partial V}=-N/(\rho_0 V^2)$.
The equilibrium pressure is hence given by (noting $\mathcal{F}=fV$)
\begin{equation}
  P=-\frac{\partial\mathcal{F}}{\partial V}
  =-f-V\frac{\partial f}{\partial \bar{n}}\frac{\partial \bar{n}}{\partial V}
  =-f+(\bar{n}+1)\frac{\partial f}{\partial \bar{n}}. \label{P_df}
\end{equation}

For a solid phase, numerical solution of the full PFC model is required to calculate this
pressure $P$ through $f(\bar n)$ of the equilibrium state. For uniform vapor or liquid phase,
we can obtain the analytic expression of $P$ based on Eq.~(\ref{V-L_Eq1}) for $f$, i.e.,
\begin{eqnarray}
  P_{\rm uniform}&=&-B_0-\frac{1}{2}C_0 \left (\bar n^2+2\bar n \right ) \nonumber \\
  && -\frac{1}{6}D_0 \left (2\bar n^3+3\bar n^2 \right )
  -\frac{1}{24}E_0 \left (3\bar n^4+4\bar n^3 \right ).
  \label{P_unif}
\end{eqnarray}
The saturation vapor pressure $P_{\rm u}^{\rm (SVP)}$ at vapor-liquid coexistence can be calculated
by substituting Eq.~(\ref{ncoexist}) for coexistence density into Eq.~(\ref{P_unif}). Some
results are depicted in Fig.~\ref{fig:Fig8}, showing the important role of $B_0$ on the
temperature dependence of $P$. When $B_0=0$ and using values given in Table~\ref{tab:table1}
for other parameters (bottom blue line in Fig.~\ref{fig:Fig8}) ${P}_{\rm u}^{\rm (SVP)}$ decreases
with the increase of temperature $\Delta T$, a behavior that is not correct. The correct
temperature-increasing behavior of ${P}_{\rm u}^{\rm (SVP)}$ is obtained only when $B_0$ is set to
be temperature dependent, such as $B_0=-4.5-3\Delta T$ given in Table~\ref{tab:table1} (upper
red line in Fig.~\ref{fig:Fig8}).

\begin{figure*}
  \centerline{\includegraphics[width=0.75\textwidth]{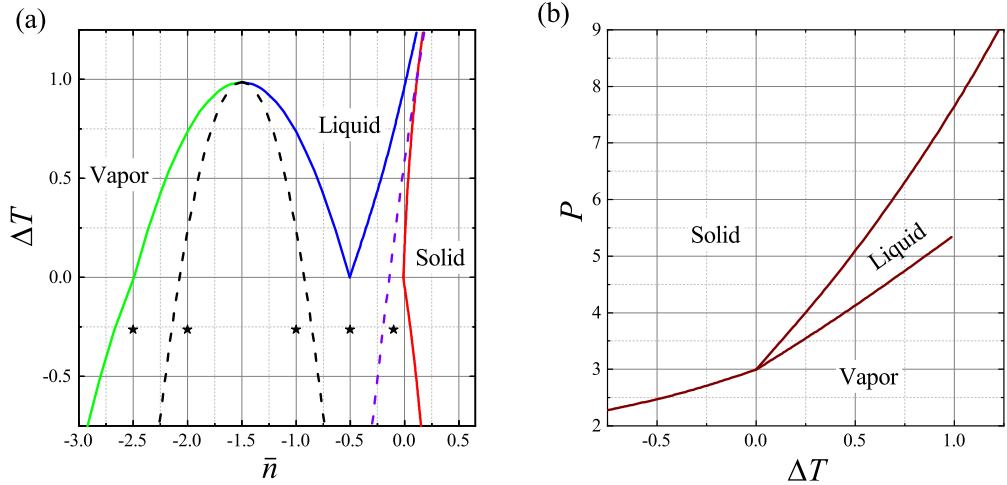}}
  \caption{Vapor-liquid-solid phase diagrams calculated numerically from the full PFC model,
    where the solid state is of 2D hexagonal phase. (a) Temperature-density phase diagram, where
    the phase boundaries for vapor, liquid, and solid states are plotted in green, blue, and
    red, respectively, the vapor-liquid spinodal is plotted as the black dashed curve, and
    the linear instability of the homogeneous state is indicated by the purple dashed line.
    Star symbols refer to the 5 parameter locations used in our numerical simulations shown
    in Fig.~\ref{fig:Fig11}. (b) Temperature-pressure phase diagram. The model parameters
    listed in Table~\ref{tab:table1} are used.}
\label{fig:Fig9}
\end{figure*}

Based on these information of $f$ and $P$, we compute the vapor-liquid-solid phase diagrams
of the full PFC model using the parameters listed in Table~\ref{tab:table1}, as shown in
Fig.~\ref{fig:Fig9}. Following the procedure described above, at each temperature $\Delta T$
the equilibrium free energy density $f$ for solid phase is evaluated numerically from simulations
of single unit cell for different values of $\bar n$ (see some examples in Fig.~\ref{fig:Fig7}).
The resulting $f$-$\bar n$ relations are then used in the common tangent construction described
in Appendix \ref{sec:Params} to obtain the temperature-density phase diagram presented in
Fig.~\ref{fig:Fig9}(a). At the same time the corresponding pressure value at phase coexistence
densities can be determined from Eqs.~(\ref{P_df}) and (\ref{P_unif}) for each $\Delta T$,
giving the temperature-pressure phase diagram in Fig.~\ref{fig:Fig9}(b).

These calculated phase diagrams possess expected properties of vapor-liquid-solid transitions
and coexistence. For example, for vapor-solid coexistence at low temperatures (with $\Delta T<0$) 
the vapor coexistence density increases with temperature but the solid one decreases, while
for liquid-solid coexistence at intermediate and high temperatures ($\Delta T>0$), both liquid
and solid coexistence densities increase with temperature [see Fig.~\ref{fig:Fig9}(a)].
In addition, both the triple point and critical point are obtained in the temperature-pressure
phase diagram [Fig.~\ref{fig:Fig9}(b)]. We have conducted some numerical simulations to verify
the phase behavior identified here, with sample results given below in Sec.~\ref{sec:Dynamics}.
All these results are consistent with experimental phase diagrams of pure materials (e.g., argon)
and those of previous computer simulations (such as the Lennard-Jones system).

\subsection{Equilibrium lattice spacing: Effects of thermal expansion and pressure}

A drawback of the previous PFC models is the lack of lattice thermal expansion effect, and
also the lack of a study of effects of pressure $P$ and density $\bar n$ on the lattice constant.
For example, in the original PFC model the equilibrium wave number in one-mode approximation is
given by $q_{\rm eq}=\sqrt{{C_2}/{2C_4}}$, which is independent of $\bar n$, $P$, and temperature
as $C_2$ and $C_4$ were assumed to be temperature independent constants \cite{prb064107}. In this
model $C_2$ and $C_4$ are set to be dependent on the temperature (see Table~\ref{tab:table1}),
and with the incorporation of high-order correlations in the model (related to three- and
four-point interactions), $q_{\rm eq}$ is affected by $\bar n$ and hence pressure $P$ via $D_{11}$
and $E_{1122}$ terms [see e.g., Eq.~(\ref{VLS_Eq2})]. Therefore, both thermal expansion and
pressure effects have been incorporated in this three-phase PFC model.

\begin{figure}
  \centerline{\includegraphics[width=0.5\textwidth]{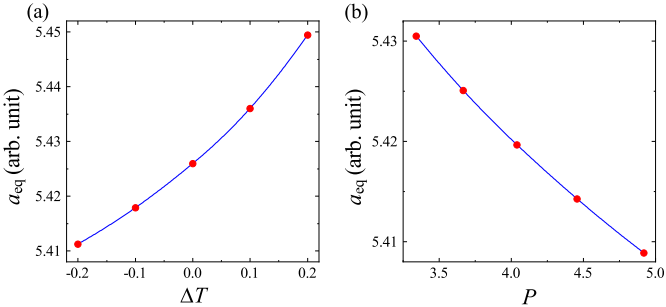}}
  \caption{Thermal and pressure effects on lattice constant.
    (a) Thermal expansion of equilibrium lattice spacing $a_{\rm eq} = 2\pi/q_{\rm eq}$ 
    under a constant pressure $P=3.87$. (b) The pressure-induced variation of $a_{\rm eq}$
    under a constant temperature $\Delta T=0$. Parameters in Table~\ref{tab:table1} for
    2D hexagonal phase are used in numerical calculations.}
\label{fig:Fig10}
\end{figure}

Results of numerical calculations of equilibrium lattice spacing $a_{\rm eq}$ for 2D hexagonal
structure are shown in Fig.~\ref{fig:Fig10}, subjected to variations of temperature and pressure.
During the process of equilibrium free energy density and phase diagram calculations described
above, $q_{\rm eq}$ has already been determined through free energy minimization at each $\bar{n}$
and $\Delta T$, yielding the corresponding lattice constant $a_{\rm eq} = 2\pi/q_{\rm eq}$. For each
value of average density $\bar{n}$, pressure $P$ is calculated numerically based on Eq.~(\ref{P_df}).
Figure \ref{fig:Fig10}(a) shows that at a constant value of $P$, $a_{\rm eq}$ increases with larger
temperature $\Delta T$, indicating a behavior with positive thermal expansion coefficient as in
a majority of materials. In addition, when temperature $\Delta T$ is kept constant while $P$ is
varied, Fig.~\ref{fig:Fig10}(b) shows that $a_{\rm eq}$ decreases with increasing $P$, consistent
with the compression effect of pressure on the lattice.

\subsection{Dynamical simulations}
\label{sec:Dynamics}

\begin{figure}
  \centerline{\includegraphics[width=0.5\textwidth]{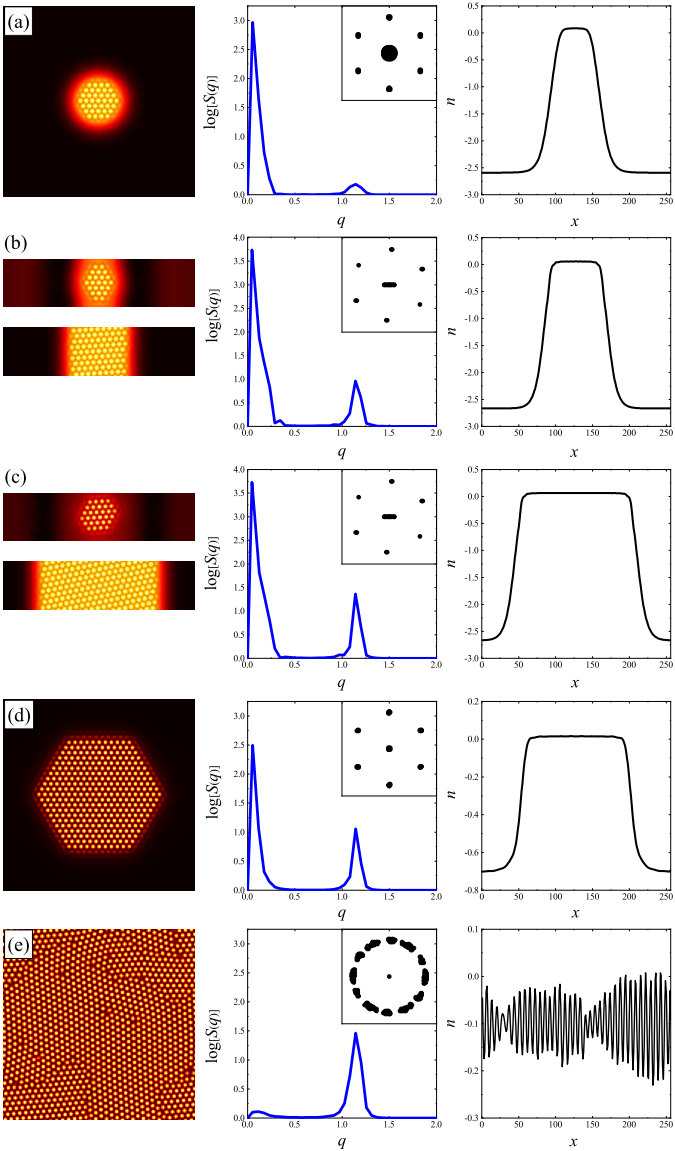}}
  \caption{Sample results from PFC simulations at $\Delta T=-0.264$, corresponding to the locations
    marked in the phase diagram of Fig.~\ref{fig:Fig9}(a) below the triple point temperature. They
    include five characteristic cases for which the initial configuration is set up as a crystalline
    nucleus of $\bar n = 0.1$ and grid size $48 \times 48$ embedded in a uniform phase of
    (a) $\bar n = -2.5$, (b) $\bar n = -2$, (c) $\bar n = -1$, and (d) $\bar n = -0.5$, or by
    (e) a random initial condition of $\bar n = -0.1$ in the whole system.
    Left column: Spatial structure at the steady state of phase coexistence [(a)--(d)] or at the
    late stage of the polycrystalline state (e). An early-stage configuration having liquid-vapor
    phase separation is also shown in (b) and (c) for initial $\bar n$ within the spinodal.
    The brighter (darker) regions correspond to higher (lower) values of density field $n$.
    Middle column: The corresponding circularly averaged structure factor, with the diffraction
    pattern shown as inset. Right column: The $y$-averaged density of the final state across the
    grid points along the $x$ direction.
  }  
\label{fig:Fig11}
\end{figure}

\begin{figure}
  \centerline{\includegraphics[width=0.5\textwidth]{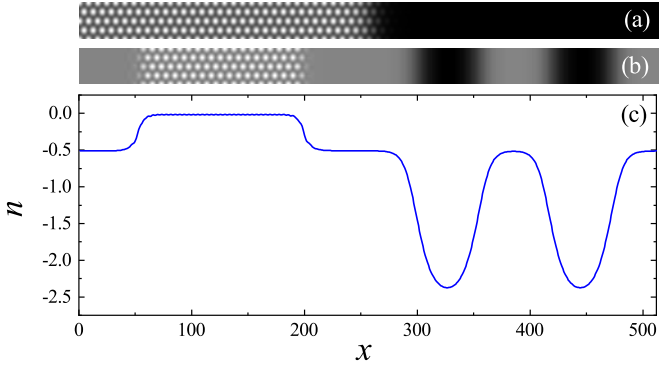}}
  \caption{PFC simulation results of vapor-liquid-solid coexistence at $\Delta T=0$, using 
    the model parameters given in Table \ref{tab:table1}. 
    (a) The initial state with half of system in solid (with $\bar{n}=-0.012$) and the 
    other half in a uniform state (with $\bar{n}=-1.5$). (b) The time-evolving state with
    vapor-liquid-solid coexistence. (c) The corresponding $y$-averaged density along the
    $x$ direction.}
\label{fig:Fig12}
\end{figure}

We have conducted dynamical simulations based on Eq.~(\ref{DynamEq}) using the model parameters
listed in Table \ref{tab:table1}, to examine the above results of vapor-liquid-solid transitions
and coexistence. Our focus is on the regime involving vapor phase at and below the triple point
temperature in the phase diagram (i.e., $\Delta T \leq 0$), with some sample results presented
in Figs.~\ref{fig:Fig11} and \ref{fig:Fig12}.

Figure \ref{fig:Fig11} shows five typical scenarios of structural evolution and phase coexistence,
with the corresponding parameter values of the initial uniform phase indicated in the
temperature-density phase diagram of Fig.~\ref{fig:Fig9}(a) at $\Delta T=-0.264$. The simulations
in Figs.~\ref{fig:Fig11}(a)--\ref{fig:Fig11}(d) were initialized from a spatially homogeneous
state of density $\bar{n}_{\rm uniform}$, with a solid seed of 2D hexagonal structure of
$\bar{n}=0.1$ placed at the center, while a random initial condition of $\bar{n}=-0.1$ was
set in the whole system of Fig.~\ref{fig:Fig11}(e). In addition to the spatial structure
configurations, the circularly averaged structure factor, diffraction pattern, and the
$y$-averaged density profile along the $x$ direction are presented in Fig.~\ref{fig:Fig11}
for the final state of simulations. The final state in Figs.~\ref{fig:Fig11}(a)--\ref{fig:Fig11}(d)
corresponds to equilibrium two-phase coexistence, for which the structure factor shows
two peaks, one at small wave number $q$ as caused by the vapor or liquid region while
the other corresponding to the hexagonal lattice inside the solid grain. For the polycrystalline
state in Fig.~\ref{fig:Fig11}(e), the small-$q$ peak of the structure factor can be attributed
to the existence of multiple grains in the sample.

When the initial value of average density $\bar n$ locates between the vapor phase boundary
and the spinodal curve of the phase diagram, such as $\bar{n}_{\rm uniform}=-2.5$ in the case of
Fig.~\ref{fig:Fig11}(a), no vapor-liquid separation occurs in the initially homogeneous
region and the system equilibrium state is characterized by the coexistence between
vapor phase and a stabilized faceted solid grain, as expected. For Fig.~\ref{fig:Fig11}(b)
with $\bar{n}_{\rm uniform}=-2.0$ and Fig.~\ref{fig:Fig11}(c) with
$\bar{n}_{\rm uniform}=-1.0$ that are close to two opposite sides of the spinodal boundary,
similar results of vapor-solid state are obtained, although with larger equilibrium solid
region appearing in the latter case, consistent with the lever rule. Both values of
$\bar{n}_{\rm uniform}$ locate within the spinodal regime, so that vapor-liquid phase separation
occurs spontaneously at the early stage of system evolution, as can be seen in the top panels
of the left column in Figs.~\ref{fig:Fig11}(b) and \ref{fig:Fig11}(c). The liquid-phase region
shrinks with time and eventually disappears, while the initial solid grain grows and saturates,
leading to an equilibrium state with vapor-solid coexistence as shown in the bottom panels
of the left column (see also the averaged density profile given in the right column).
When initially $\bar{n}_{\rm uniform}=-0.5$, lying between the spinodal curve and linear
instability line, the final equilibrium state shows a coexistence between liquid region
and the embedded faceted solid grain, as seen in Fig.~\ref{fig:Fig11}(d). Finally, at
$\bar{n}=-0.1$ which is beyond the linear instability line, the initial homogeneous
state is linearly unstable, leading to the spontaneous formation of the crystalline
structure across the system which evolves to a polycrystalline configuration shown in
Fig.~\ref{fig:Fig11}(e). The system consists of various topological defects including
dislocations and grain boundaries, resulting in the spatial oscillations of the $y$-averaged
density profile across the $x$ direction as presented in the bottom-right panel of the figure.

To further illustrate the phenomenon of vapor-liquid-solid coexistence, we simulate a
system of 2D slab configuration at $\Delta T=0$, starting with half of the slab occupied
by solid phase with $\bar{n}=-0.012$ while the other half by a homogeneous state with
$\bar{n}=-1.5$ (at the middle of the spinodal regime), as shown in Fig.~\ref{fig:Fig12}(a).
Through spinodal decomposition, the system spontaneously evolves into a mixture of vapor,
liquid, and solid [Fig.~\ref{fig:Fig12}(b)]. The resulting smoothed average density
[Fig.~\ref{fig:Fig12}(c)] closely matches to that of the equilibrium phase diagram in
Fig.~\ref{fig:Fig9}(a), i.e., $\bar n =-2.493$, $-0.507$, and $-0.012$ for vapor, liquid,
and solid phases, respectively. It is noted that in this simulated system the solid region
is actually of higher energy due to the existing of interfaces and nonzero interfacial energy,
and thus shrinks slowly with time during the system evolution.

\section{Conclusions}
\label{sec:summary}

We have introduced a new PFC model with high-order correlations featuring three- and
four-body interactions to examine the transitions and coexistence among vapor, liquid, and
crystalline solid phases within a single continuum density-field description,
without making any pre-assumptions other than the basic requirement of system rotational
invariance. The advantage of this model has been demonstrated in terms of its
simple form, genericness and flexibility of parameter choices in achieving three-phase coexistence
and transitions, including both vapor-liquid-solid and the unusual vapor-solid-liquid transition
sequence. Through both theoretical analysis and numerical computation, the conditions of phase
coexistence are identified, as well as temperature-density and temperature-pressure phase
diagrams incorporating vapor-liquid-solid triple point and vapor-liquid critical point, which
qualitatively agree with the well-known results of previous experiments and atomistic simulations.
Various scenarios of vapor-solid, liquid-solid, and vapor-liquid-solid coexistence, vapor-liquid
phase separation, and structural evolution are verified through dynamical simulations of the model.

In addition, several material properties missing in the previous PFC models, including temperature
dependence of saturation vapor pressure, thermal expansion, and compression effect of pressure
on lattice constant, can be well produced in this model, with outcomes consistent with the
known results. Thus the approach developed here, which well describes the vapor-liquid-solid
phase behaviors and the corresponding material properties, can serve as a valuable tool for
modeling the material growth and evolution processes including both vapor- and liquid-based
and vapor-liquid-solid growth.

\appendix
\section{Structure factor of the homogeneous state}
\label{sec:Sq}

The structure factor for homogeneous fluids can be determined from a linear analysis of the
dynamic equation governing the density field $n$. In the homogeneous fluid state the density
can be decomposed as $n(\mathbf{r},t)=\bar{n}+\delta n(\mathbf{r},t)$ with a small fluctuation
$\delta n(\mathbf{r},t)$. Linearizing the dynamical Eq.~(\ref{DynamEq}), in Fourier space we get
\begin{equation}
  \frac{\partial \delta \hat{n}_{\mathbf{q}}}{\partial t} 
  =-\gamma_{\mathbf{q}} \delta \hat{n}_{\mathbf{q}} + \hat{\eta}_{\mathbf{q}},
  \label{DynamEq_nk}
\end{equation}
where $\delta \hat{n}_{\mathbf{q}}$ is the Fourier transform of $\delta n$, and
\begin{eqnarray}
  \gamma_{\mathbf{q}} &=& {q}^2 \left[ -\left ( {C_{0}}+{\bar{n}}{{D}_{0}}
    +\frac{1}{2}\bar{n}^{2}E_0 \right ) + \left ( C_{2}+\frac{2}{3}{\bar{n}}{D_{11}} \right)
    {q}^{2} \right. \nonumber\\
    && \left. -\left( C_{4}+\frac{1}{12}\bar{n}^{2}E_{1122} \right) {q}^{4}
    +C_{6} {q}^{6} \right ].
  \label{Gamma_k}
\end{eqnarray}
Here we have introduced a noise term $\hat{\eta}_{\mathbf{q}}$, which is the Fourier component
of the noise field $\eta$ satisfying $\langle \eta(\mathbf{r},t) \rangle =0$ and
$\langle \eta(\mathbf{r},t)\eta(\mathbf{r}',t') \rangle = - 2k_BT \nabla^2
\delta(\mathbf{r}-\mathbf{r}') \delta(t-t')$; thus
$\langle \hat{\eta}_{\mathbf{q}}(t)\hat{\eta}^*_{\mathbf{q}'}(t') \rangle
= 2k_BT q^2 \delta(\mathbf{q}-\mathbf{q}') \delta(t-t')$.

Following the procedure given in Ref.~\cite{pre51605}, we obtain the solution of
Eq.~(\ref{DynamEq_nk}) as
\begin{equation}
  \delta \hat{n}_{\mathbf{q}}(t)=e^{-\gamma_{\mathbf{q}}t} \delta \hat{n}_{\mathbf{q}}(0)
  +e^{-\gamma_{\mathbf{q}}t} \int_{0}^{t} ds e^{\gamma_{\mathbf{q}}s} \hat{\eta}_{\mathbf{q}}(s),
  \label{solution_nk}
\end{equation}
and calculate the fluid-state structure factor by
\begin{eqnarray}
  S(\mathbf{q},t) &=& \left \langle \left| \delta \hat{n}_{\mathbf{q}} \right|^2 \right \rangle
  =\left \langle \left| \delta \hat{n}_{\mathbf{q}} \delta \hat{n}_{-\mathbf{q}} \right| \right \rangle
  \nonumber \\
  &=& e^{-2\gamma_{\mathbf{q}}t}\left \langle \left| \delta \hat{n}_{\mathbf{q}}(0) \right|^2 \right \rangle
  \nonumber\\
  && +e^{-2\gamma_{\mathbf{q}}t} \int_0^t {dsds' e^{\gamma_{\mathbf{q}}(s+s')} \left \langle \hat{\eta}_{\mathbf{q}}(s)
    \hat{\eta}_{-{\mathbf{q}}}(s') \right \rangle} \nonumber \\
  &=&e^{-2\gamma_{\mathbf{q}}t} S({\mathbf{q}},0) +\frac{k_BT q^2}{\gamma_{\mathbf{q}}}
  \left( 1-e^{-2\gamma_{\mathbf{q}}t} \right). \label{Sq}
\end{eqnarray}
The equilibrium fluid-state structure factor is defined by $S(\mathbf{q})=S(\mathbf{q},t \to \infty)$,
yielding
\begin{equation}
  S(q)=\frac{k_BTq^2}{\gamma_{\mathbf{q}}}
  =\frac{k_BT}{-\left(C_{0}'-C_{2}'q^2+C_{4}'q^4-C_{6}'q^6 \right)},
\label{Sq_liquid}
\end{equation}
where
\begin{eqnarray}
  C_0' = C_0 + \bar{n} D_0 + \frac{1}{2}\bar{n}^2E_0, \qquad
  && C_2' = C_2 + \frac{2}{3}\bar{n}D_{11}, \nonumber\\
  C_4' = C_4 + \frac{1}{12} \bar{n}^2 E_{1122}, \qquad && C_6' = C_6.
  \label{Para_trans}
\end{eqnarray}

\section{Phase coexistence and model parameters selection}
\label{sec:Params}

In this appendix we show the procedure of identifying three-phase coexistence in this PFC model
and how to choose the corresponding model parameters. According to the common tangent rule,
in the equilibrium state the coexistence between any two phases is determined by equal chemical
potential $\mu_n$ and equal pressure $P_n$, i.e.,
\begin{eqnarray}
  && \left. \frac{\partial f}{\partial \bar n} \right |_1
  = \left. \frac{\partial f}{\partial \bar n} \right |_2 = \mu_n \nonumber \\
  && f_1 - \mu_n {\bar n}_1 = f_2 - \mu_n {\bar n}_2 = -P_n, \label{common_tangent}
\end{eqnarray}
giving coexistence densities ${\bar n}_1$ and ${\bar n}_2$ for phase 1 and 2, respectively.
[Note that $P_n = P-\mu_n$ if compared to Eq.~(\ref{P_df}) for pressure $P$ obtained through
density $\bar{\rho}$.]

In this PFC model the free energy density for uniform phase, either vapor ($f_{\rm vapor}$)
or liquid ($f_{\rm liquid}$), is known from Eq.~(\ref{V-L_Eq1}), and the exact solution of
coexistence densities $\bar{n}_{\rm vapor}$ and $\bar{n}_{\rm liquid}$ is given by
Eq.~(\ref{ncoexist}). Thus from Eq.~(\ref{common_tangent})
we can get the value of $\mu_n = (f_{\rm liquid}-f_{\rm vapor})/(\bar{n}_{\rm liquid}-\bar{n}_{\rm vapor})$
from vapor-liquid coexistence, and the following conditions governing the vapor-liquid-solid
coexistence
\begin{eqnarray}
  && \left. \frac{\partial{f_{\rm solid}}}{{\partial \bar{n}}} \right |_{\bar{n}_{\rm solid}} = \mu_n
  =\frac{f_{\rm liquid}-f_{\rm vapor}}{\bar{n}_{\rm liquid}-\bar{n}_{\rm vapor}}, \label{solid1}\\
  && f_{\rm solid}=f_{\rm liquid}+\frac{(\bar{n}_{\rm solid}
    -\bar{n}_{\rm liquid})(f_{\rm liquid}-f_{\rm vapor})}{\bar{n}_{\rm liquid}-\bar{n}_{\rm vapor}}.
  \label{solid2}
\end{eqnarray}
For the full model, the solid-state free energy density $f_{\rm solid}$ is calculated from
the steady state of the numerical solution of dynamical Eq.~(\ref{DynamEq}) for a single-crystal
unit cell (see Sec.~\ref{sec:phasediagram}), while phase coexistence is determined by the common
tangent construction described above, given the known results of coexisting vapor and liquid
phases in Eqs.~(\ref{V-L_Eq1}) and (\ref{ncoexist}).

There are many adjustable parameters in the model, including $B_0$, $C_{j=0,2,4,6}$, $D_0$,
$E_0$, $D_{11}$, and $E_{1122}$. For simplicity, we first fix the values of $B_0$, $C_0$,
$D_0$, and $E_0$ so that the properties of vapor and liquid phases are pre-determined, such as
the coexistence densities $\bar{n}_{\rm vapor}$ and $\bar{n}_{\rm liquid}$, the resulting $f_{\rm vapor}$
and $f_{\rm liquid}$, and saturation vapor pressure (see Fig.~\ref{fig:Fig8}). Value of $C_6$ is
also chosen in advance based on its effect on high-order modes (see e.g., Fig.~\ref{fig:Fig5}).
We then have only four parameters $C_2$, $C_4$, $D_{11}$, and $E_{1122}$ left to be determined,
to satisfy the conditions of three-phase coexistence.

We first follow this procedure with the use of one-mode approximation for solid phase
to determine all the model parameters, and then slightly adjust them (to account for the
discrepancy between one-mode and full-model results) to obtain phase coexistence and phase
diagrams from numerical calculations of the full PFC model. The one-mode free energy density
$f_{\rm solid}$ is given by Eq.~(\ref{F_Stripes}) for 1D stripe and by Eq.~(\ref{F_Rods}) for
2D hexagonal phase, respectively. Its equilibrium state with minimum free energy is obtained
by solving
\begin{equation}
\frac{\partial{f_{\rm solid}}}{{\partial A}}=0, \qquad \frac{\partial{f_{\rm solid}}}{{\partial q}}=0.
\label{solid3}
\end{equation}
We thus have four equations in Eqs.~(\ref{solid1})--(\ref{solid3}) to be solved numerically
for seven unknown variables $C_2$, $C_4$, $D_{11}$, $E_{1122}$, $A$, $q$, and $\bar{n}_{\rm solid}$
in one-mode approximation. To identify the allowed values of model parameters $C_2$, $C_4$,
$D_{11}$, and $E_{1122}$ yielding three-phase coexistence, we solve these equations under
specific values of $A$, $\bar{n}_{\rm solid}$, and $q$ (fixed as $2/\sqrt{3}$ here in 2D one-mode
expansion at the triple point) that can lead to the existence of solution. The corresponding
results are presented in Fig.~\ref{fig:Fig6}.

\begin{acknowledgments}

Z.-L.W. acknowledges support from the China Postdoctoral Science Foundation (Grant No.~2020M670275).
Z.R.L. acknowledges support from the National Natural Science Foundation of China (Grant No.~21773002). 
Z.-F.H. acknowledges support from the U.S. National Science Foundation under Grant No.~DMR-1609625.
W.D. acknowledges support from the National Natural Science Foundation of China (Grant No.~51788104),
the Ministry of Science and Technology of China (Grant No.~2016YFA0301001), and the Beijing Advanced
Innovation Center for Materials Genome Engineering.

\end{acknowledgments}

\bibliography{VLS_pfc_references}

\end{document}